# ADAPTIVE SLOT ALLOCATION AND BANDWIDTH SHARING FOR PRIORITIZED HANDOFF CALLS IN MOBILE NETWOKS

**S.Malathy**
*Research Scholar*
*Anna University*
*Coimbatore*
joymalathy@gmail.com

**G.Sudhasadasivam**
*Professor, CSE Department*
*PSG College of Technology*
*Coimbatore*

**K.Murugan**
*Lecturer, IT Dept*
*Hindusthan Institute of Tech*
*Coimbatore*

**S.Lokesh**
*Lecturer, CSE Dept*
*Hindusthan Institute of Tech*
*Coimbatore*

*Abstract* - *Mobility management and bandwidth management are two major research issues in a cellular mobile network. Mobility management consists of two basic components: location management and handoff management. To Provide QoS to the users Handoff is a key element in wireless cellular networks. It is often initiated either by crossing a cell boundary or by deterioration in the quality of signal in the current channel. In this paper, a new admission control policy for cellular mobile network is being proposed. Two important QoS parameter in cellular networks are Call Dropping Probability (CDP) and Handoff Dropping Probability (HDP). CDP represents the probability that a call is dropped due to a handoff failure. HDP represents the probability of a handoff failure due to insufficient available resources in the target cell. Most of the algorithms try to limit the HDP to some target maximum but not CDP. In this paper, we show that when HDP is controlled, the CDP is also controlled to a minimum extent while maintaining lower blocking rates for new calls in the system.*

*Index Terms*— **Wireless Cellular Networks, Handoff Dropping Probability, Call Dropping Probability, Resource Allocation, Prioritization Schemes.**

## 1. INTRODUCTION

Due to the increased urge to use the wireless communication in a satisfied way, a promised Quality of Service (QoS) is required to manage the incoming new calls and handoff calls more efficiently. The Geographical area is divided into smaller areas in the share of hexagon. These hexagonal areas are called as cells. A Base Station (BS) is located at each cell. The Mobile Terminals (MT) within that region is served by these BS. Before a mobile user can communicate with other mobile user in the network, a group of channels should be assigned. The cell size plays a major role in the channel utilization. A user has to cross several cells during the ongoing conversation, if the cell size is small. During the ongoing conversation, the call has to be transferred from one cell to another to achieve the call continuation during boundary crossing.

Here comes the role of handoff. Transferring the active call from one cell to another without disturbing the call is called as the process of Handoff. Hand is otherwise a "make before break" process. Time slot, Frequency band, or code word to a new base station [1] may be the terms of call transfer from a cell to another.

A typical Cellular network is shown in figure 1. A limited frequency spectrum is allocated. But it is very successfully utilized because of the frequency reuse concept. To avoid the interference while neighboring cells are utilizing the same frequency, the group of channels assigned to one cell should be different from the neighboring cells. The MTSO scans the residence of the MS and assigns the channel to that cell for the call.

If the MS is travelling while the call is in progress, the MS need to get a new channel from the neighboring BS to continue the call without dropping. The MSs located in the cell share the available channels. The Multiple Access Methods and channel allocation schemes govern the sharing and allocating the channels in a cell, respectively.

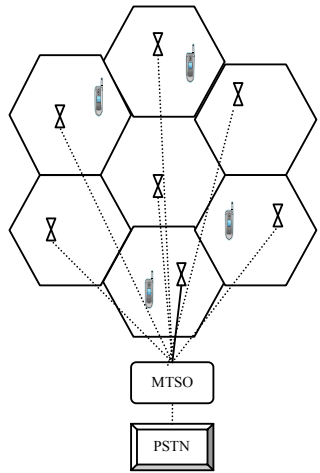

FIGURE 1 CELLULAR NETWORK

The Scenario of a basic cellular network is depicted in Figure1.

The resource management in the cellular system deals with CAC, Utilization of Power and channel allocation strategy. The channel allocation strategy may be Fixed or Dynamic. The resource allocation is shown in Figure 2.

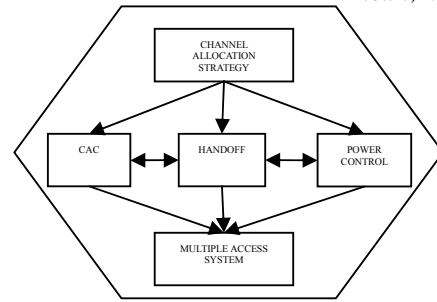

FIGURE 2 RESOURCE MANAGEMENT IN CELLULAR NETWORKS

Call Admission Control denotes the process of admitting a fresh call or a handoff call based on the availability of resources.

## II LITERATURE SURVEY

Various handoff schemes proposed [2] are Guard channel scheme (GCS), Handoff based on Relative Signal Strength [4], Handoff based on Relative Signal Strength with threshold, Handoff based on Relative Signal Strength with Hysteresis and threshold [3], Handoff based on Prediction techniques [5]. When MS moves from one cell to another, the corresponding BS hands off the MSs Call to the neighbor. This process is done under the control of MTSO. The handoff in initiated based on various parameters like signal strength received from BS, travelling speed of the MS etc.

A handoff method based on the kinds of state information [6] that have been defined for MSs, as well as the kinds of network entities that maintain the state information has been devised. The handoff decision may be made at the MS or network. Based on the decision, three types of handoff may exist namely, Network-Controlled Handoff, Mobile-Assisted Handoff, and Mobile-Controlled Handoff [7]. Handoff based on Queuing is analyzed [7] for voice calls. The Queue accommodates both the originating calls and handoff requests [9]. Handoff schemes with two-level priority [10] have been proposed. How the non-real-time service has to be incorporated and its effect needs to be taken into consideration is proposed [11]. A new two-dimensional model for cellular mobile systems with pre-emptive priority to real time service calls [12] is proposed. In [13] the concept of prioritization of handoff calls over new calls since it is desirable to complete an ongoing call rather than accepting a new one is employed.

In [14], a situation where the handoff calls are queued and no new calls are handled before the handoff calls in the queue is presented. By combing guard channel and queue schemes performs better [15]. [16] developed a non-preemptive prioritization scheme for access control in cellular networks.

## III. SYSTEM DESCRIPTION

If users request connection to the base station at the same time, the system checks the type of origin of the call. The handoff decision may be made by the MS or the network based on the RSS, Traffic pattern, Location management etc., while handoff is made the channel assignment plays an important role. The total channels in the BS can be allocated to different types of calls. If the originating calls and handoff calls are treated in the same way, then the request from both kinds are not served if there are no free channels.

In another scheme, Priority is given to the handoff call request by reserving a minimum number of channels to the handoff call. If there is N number of channels available, the R number of channels is reserved to the handoff calls and the remaining (N-R) channels are shared by the handoff and originating call requests. The handoff call request is dropped only if there are no channels available in the cells. To overcome this drawback of dropping the handoff calls, our system proposes an new model of queuing scheme in with the handoff calls and originating calls are queued to get the service. The priority is more for the handoff calls than the originating calls.

The following assumptions are made over the calls.
   a) The arrival pattern of the calls follows the Poisson process.
   b) The cell consists of N Channels. If free channels exist, both the calls will be served. If channels are not available, then the originating calls will be dropped.
   c) Priority is given to the handoff calls on based on the call dwell time in the cells. The priority is low for a longer dwell time calls than the shorter calls. The channel holding time is assumed to have exponential distribution.

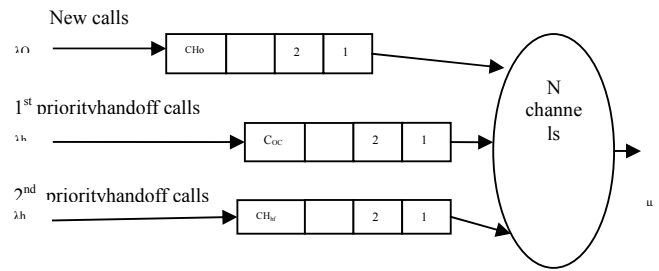

FIGURE 3 QUEUEING CALLS

   d) Two types of Queues are assumed. The queue for handoff calls $Q_{HC}$ and queue for originating calls $Q_{OC}$ respectively.
   e) If no channels are available the handoff calls are queued in $Q_{HC}$, whose capacity is $C_{HC}$. The originating calls are queued in $Q_{OC}$, only if the available channels at the time of arrival are less than (N-R). The originating call is blocked if the queue is full.
   f) Queue is cleared if the call is completed or the user moves away from the cell.

g) The capacity $C_{HC}$ of $Q_{HC}$ is large enough so that blocking probability of the handoff call is neglected.

The channel holding time TH can be calculated by using the following formula

$$\int_0^\infty e^{-\mu Ht} dt = \int_0^\infty ((1 - \frac{\lambda_n}{\lambda} F_{TH_n}(t) - \frac{\lambda_n}{\lambda} F_{TH_h}(t))dt \qquad (1)$$

where $F_{THn}(t)$ and $F_{THh}(t)$ are actual distribution of channel holding time for new and handoff calls. [17]

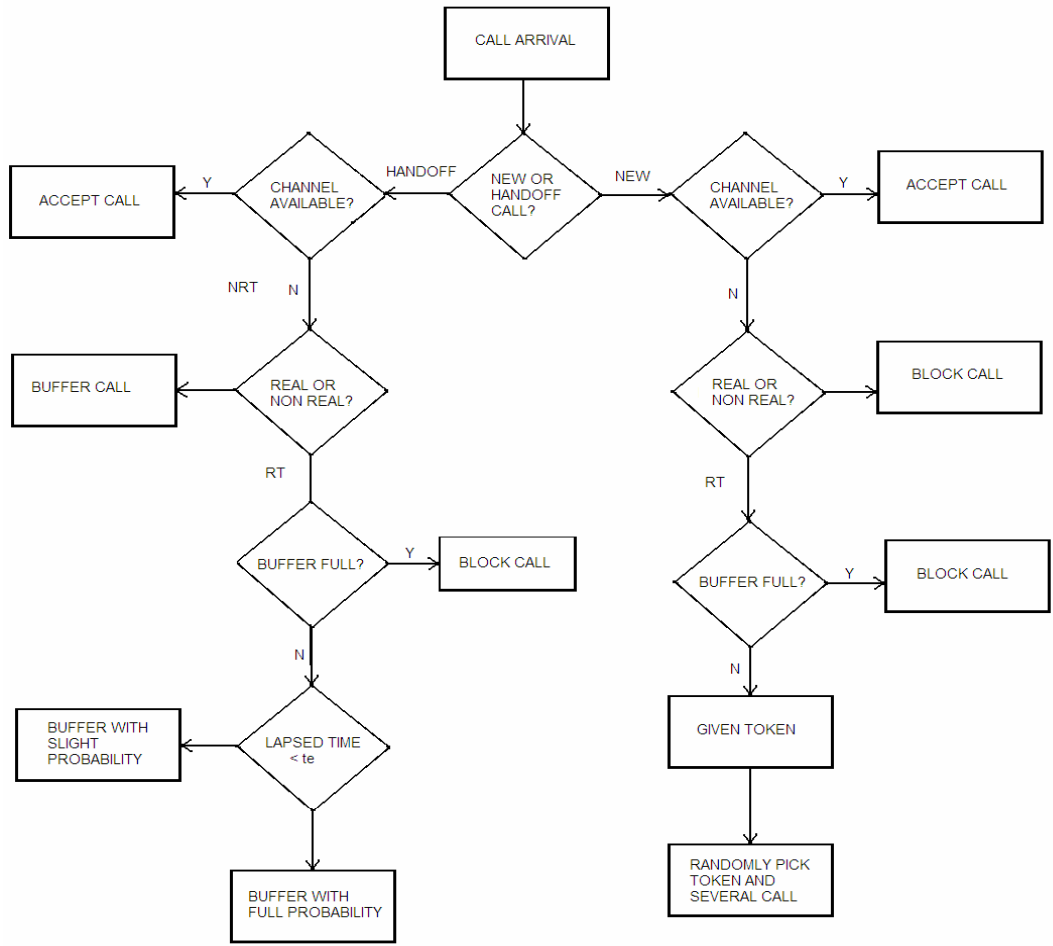

FIGURE 3 CHANNEL ALLOCATION ALGORITHM

## IV RESULTS

In this paper, a dynamic sharing of channels for the handoff calls and new calls has been proposed. In the proposed scheme, when there is no channels, the reserved channels for handoff calls of real time traffic gets shared dynamically shared by handoff calls of non-real-time traffic. The comparison between a normal bandwidth reservation scheme and the proposed model is simulated. It is shown that, the call blocking probability as well as the handoff dropping probability is reduced when compared to the traditional algorithms even when traffic is increased.

TABLE 1 COMPARISON BETWEEN EXISTING & PROPOSED SCHEMES

| Parameter | Existing Scheme | Proposed Scheme |
|---|---|---|
| Channel Utilization | Full | Reduced |
| Traffic Management | Controlled | Controlled |
| Call Dropping Probability | Reduced | Reduced |
| Call Blocking Probabilty | Not Decreased | Decreased as well |

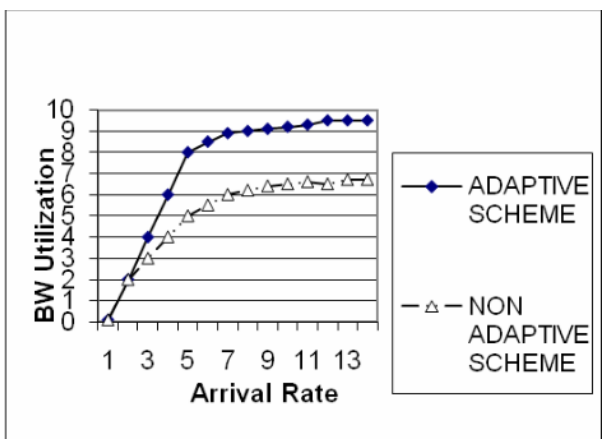

RESULT 1
BANDWIDTH UTILIZATION VERSUS CALL ARRIVAL RATE

The above graph shows that by adopting the new algorithm the bandwidth utilization is considerably increased with the increase in call rate.

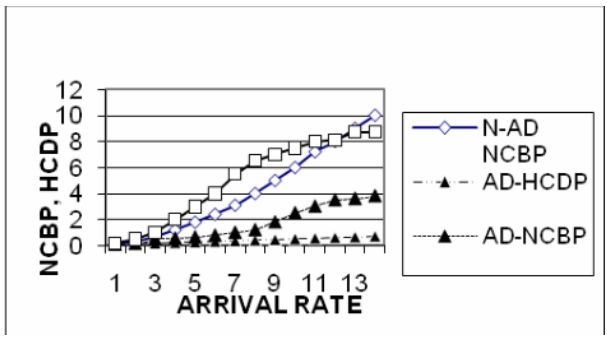

RESULT 2
CALL BLOCKING PROBABILITY VERSUS CALL ARRIVAL RATE

The New Call Blocking Probability and the Handoff Call Dropping Probability with an increase in call arrival rate in a cell is reduced when compared to the traditional algorithm.

## IV CONCLUSION

In this paper, we have showed that by integrating the concept of buffering and dwell time of the call, the New Call blocking probability and handoff call dropping probability has been considerably reduced. In future, this work can be extended for different types of calls and integrated services like data and images.

Author Profile:

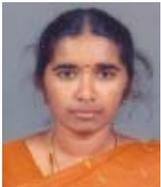 Ms.S.Malathy is currently a research Scholar in Anna University Coimbatore. She has presented nearly 10 paper in national and international conferences.Her research areas include Mobile networks and wireless communication.

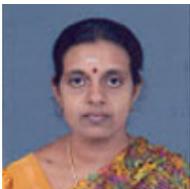 Dr *G Sudha Sadasivam* is working as a Professor in Department of Computer Science and Engineering in PSG College of Technology, India. Her areas of interest include, Distributed Systems, Distributed Object Technology, Grid and Cloud Computing. She has published 20 papers in referred journals and 32 papers in National and International Conferences. She has authored 3 books. She has coordinated two AICTE - RPS projects in Distributed and Grid Computing areas. She is also the coordinator for PSG-Yahoo Research on Grid and Cloud computing. You may contact her at sudhasadhasivam@yahoo.com

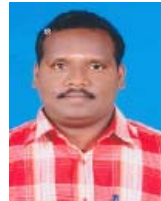 Mr.K.Murugan is currently a research Scholar in Karpagam University Coimbatore. He has a teaching experience of 15 years.He has presented various papers in national and international conferences. His research areas include Mobile networks, Grid Computing, Data Mining.

Mr.S.Lokesh is currently a research Scholar in Anna University Trichy.. He has a teaching experience of 5 years.He has presented nearly 6 papers in national and international conferences. His research areas include Mobile networks, Digital Image Processing, Signal Processing